# Dynamic Pricing of Applications in Cloud Marketplaces using Game Theory

Safiye Ghasemi[1], Mohammad Reza Meybodi[2*], Mehdi Dehghan Takht-Fooladi[2], and Amir Masoud Rahmani[1]

[1]Department of Computer Engineering, Science and Research Branch, Islamic Azad University, Tehran, Iran.
[2]Computer Engineering and Information Technology, Amirkabir University of Technology, Tehran, Iran
*mmeybodi@aut.ac.ir

**Abstract:** The competitive nature of Cloud marketplaces as new concerns in delivery of services makes the pricing policies a crucial task for firms; so that, pricing strategies has recently attracted many researchers. Since game theory can handle such competing well this concern is addressed by designing a normal-form game between providers in current research. A committee is considered in which providers register for improving their competition-based pricing policies. The functionality of game theory is applied to design dynamic pricing policies. The usage of the committee makes the game a complete-information one, in which each player is aware of every others' payoff functions. The players enhance their pricing policies to maximize their profits. The contribution of this paper is the quantitative modeling of Cloud marketplaces in form of a game to provide novel dynamic pricing strategies; the model is validated by proving the existence and the uniqueness of Nash equilibrium of the game.

## 1. Introduction

From 2007, Cloud computing has been emerged as one of the most attractive technologies in IT industry (Buyya, 2009; Hurwitz, 2010; Szabo, 2014; Zhang, 2014). An increasing number of companies are taking advantage of services provided by Cloud computing. The services are in terms of Infrastructure as a Service (IaaS), Platform as a Service (PaaS) and Software as a Service (SaaS), which are offered by different providers. IaaS providers prepare computing resources and storage resources (Hurwitz, 2010; Di Valerio, 2013) in form of virtual machines (VMs). These computing resources are requested by PaaS/SaaS providers or industrial/academic organizations for their applications to be run without necessity to maintain underlying infrastructures (Di Valerio, 2013; Truong-Huu, 2014). Nowadays, many Cloud computing providers compete with each other in maximizing their profit. The competition between SaaS providers has been emerged as a new challenge in this area. SaaS providers offer software applications and their related services (Buyya, 2009). The competition between providers may cause the market and dynamic prices be changed over time; therefore, the pricing strategies of providers must be economically efficient (Di Valerio, 2013). There are a lot of researches that focus on pricing and strategic behavior in the Internet markets and Cloud computing (Anselmi, 2014). Different pricing researches are studied in (El-Fattah, 1976; Kauffman, 2013; Nan, 2013; Narahari, 2005; Yaïche, 2000). A price optimization approach in a free competitive market is investigated in (El-Fattah, 1976); besides, the proposed approach in (Kauffman, 2013), maximizes users' profit by transferring their application to other providers or continuing using the current provider. Furthermore, a research (Nan, 2013), considered dynamic pricing mechanisms of providers with different levels of services; users select a proper based on some parameters such as response time, security and storage capacity. Communication of competitive providers and users in form of a game model is presented in (Narahari, 2005); users tend to choose the services with the best quality (QoS) while several service providers cooperate with each other in an oligopoly market for attracting more and more users and increase the profits. However, just a few works exist on competition between SaaS providers in order to increase their users. It is to be noted that users usually prefer a provider with the least price. Thus, the offered prices have an essential impact on number of users. Besides, pricing strategy has a significant influence on the profit-maximizing strategies of companies (Lehmann, 2009). Pricing of applications is computed by considering some properties, such as price formation, structure of payment flow, price discrimination, and assessment base (Narahari, 2005; Lehmann, 2009; Mathew, 2010). Most researches, which have modeled the interactions of providers or users in form of game, have mainly focused on optimal allocation of resources in Cloud providers, but none study the competition-based pricing models especially for applications.

In current paper, competitive pricing model is studied, which considers the pricing concerns of an application. We design a game of which each player is fully aware of the game and other players, known as a complete information game (Fudenberg, 1996). The players of the game are SaaS providers, who tend to attract Cloud users to maximize their final profit; they try to compute a proper price for the current request. The unique features of our work are as follows. First, this work has an analytical insight into Cloud markets and provides a quantitative modeling of these markets in form of a game between providers. Goal of the players in game is to attract users by offering proper prices; and based on the game model, the equilibrium is computed using Nash equilibrium primitives. This paper lies in trying to capture the strategic dynamics of providers as a strategic form game and the competition-based pricing model, which considers the main parameters of pricing applications. Second, the work covers some novel considerations in pricing policy of applications, which makes pricing more flexible.

The rest of this paper is organized as follows. In Section2, we discuss the preliminaries of SaaS providers' interactions in Cloud computing and introduce game theory concepts. The proposed distributed algorithm, which leads to the optimal solution for the application pricing game is formulated in Section3. The experimental results are reported in Section4. Finally, the paper is summarized with some concluding remarks in Section5.

## 2. Problem Statement and Notations

Cloud computing has a powerful paradigm in request processing for delivery of applications through provisioning of virtualized resources (Buyya, 2009). We suppose a Cloud computing marketplace which delivers application to the users; the overall architecture of the proposed Cloud market is depicted in Fig.1. Services of Cloud computing are consumed over the Internet; they can be accessed by users either via web browsers, known as a direct access or by the application programming interfaces (APIs), named as indirectly access (Stanoevska, 2009).

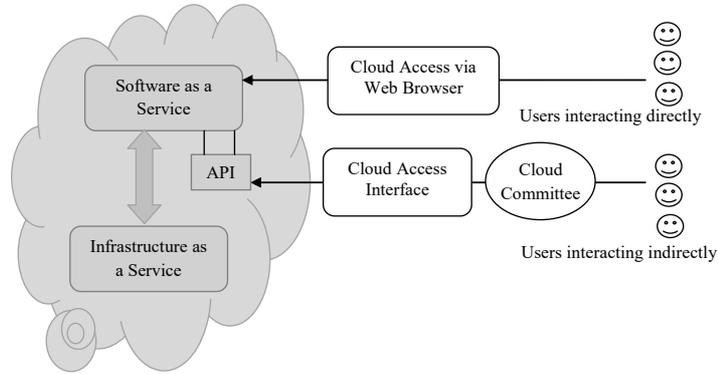

**Fig.1** interacting of Cloud computing marketplace with users via API (indirect) or via a web browser (direct)

A unit, named *Cloud Committee*, in which users can join and demand applications via the API is considered; it can be assumed as a central coordinator between users and providers. The coordinator receives users' requests; then, it transfers a request to the registered providers. After receiving the request, providers offer an optimal price for provisioning the requested application; then, users are notified of prices by the committee. Finally, users contact to a SaaS provider whose service is the best based on its offered price and the performance. As depicted in Fig.1, requests can be processed directly or indirectly. If a user demands services via the web browser of a SaaS provider (directly), the SaaS provider offers a price to the user without any comparison with other available SaaS provider. Otherwise, via indirect access, providers, who have registered in the committee, perform dynamic comparative pricing. In this case, the request is sent to all registered SaaS providers and afterwards the providers offer their prices based on a game-theoretic model; then, the best offer is chosen.

### 2.1 User requests

As mentioned previously, a user demands applications from SaaS providers via *Cloud Committee*. A request consists of parameters such as application identification, required configuration, and time period of the request. The configuration is introduced in form of parameters of VMs such as type, memory, and price. These properties comprise a VM model named VMM as {*Size*, *Memory*, *Core*, *Storage*, *HostOS*, *HourCost*}. Per unit prices are determined to charge users for using the resources according to the *Size* of each virtual resource. The parameters *Memory*, *Core*, *Storage* and *HostOS* are used for finding the proper resources to host the demanded application; *HourCost* is applied for computing the operational cost of resources in a provider.

The requests are commonly gathered in a request pool in form of a vector, named **REQ**=<$Req_1$, $Req_2$, …, $Req_q$> by *Cloud Committee*, where $Req_r$ demonstrates request $r$ in **REQ** as { $AppID_r$, $W_r$, $\tau_r$, $Pay_r$, $Prf_r$}. $AppID_r$ presents the requested application of $Req_r$, $W_r$ is the willingness of the user to pay and it is the maximum amount that the user is willing to pay; it is somehow the budgetary constraint of the user for $Req_r$; $τ_r$ represents the duration of time that the application $AppID_r$ is required; $Pay_r$ introduces the payment flow of the user which may be single payment or regularly recurring payment, and $Prf_r$ shows whether the user requires a certain performance level guarantees for a determined price or not; it is to be noted that such guarantees increase the price of the request. A more detailed discussion on the parameters is presented in Section 2.3.

**Table 1** System parameters

| Notation | Declaration |
|---|---|
| $Req_r$ | Request $r$ of the Cloud market |
| q | Number of requests sent to providers of *Cloud Committee* |
| VMM | The considered model of virtual machine |
| $τ_r$ | The duration of time the application is needed in $Req_r$ |
| $R_i$ | Number of VMs that provider $i$ bought from IaaS provider |
| $L_i$ | Number of applications that provider $i$ bought from software developers |
| N | Number of SaaS providers or players of game DPG |
| $α_i$ | Per unit benefit of virtual resources of provider $i$ |
| $β_i$ | Per unit benefit applications of provider $i$ |
| S | Possible strategies of DPG |
| $S_i$ | Possible strategies of SaaS provider $i$ (player $i$) |
| s | Strategy profile of players in DPG |
| $s_i$ | Selected strategy of player $i$ |
| $μ_i$ | Number of services of the requested application in $Req_r$ |
| $P_i$ | Offered price of SaaS provider $i$ (player $i$) for $Req_r$ |
| $c_{ij}$ | Infrastructural cost that a provider needs to pay when providing resources for $App_{ij}$ |
| $C_i$ | Cost of providing $Req_r$ for SaaS provider $i$ (player $i$) |
| $θ_{ij}$ | Price of application $j$ in provider $i$ |
| $ω_i$ | The predefined parameter by provider $i$ |
| $u_i(s)$ | Payoff function for player $i$ while playing strategy profile $s$ |

The adopted notations in this research are summarized in Table 1.

*2.2 Market state*

The received requests are delivered to the registered SaaS providers simultaneously by *Cloud Committee*. Let SaaS provider $i$ bought $R_i$ numbers of VMs of different types; for each VM $k$ that is offered by provider $i$, a per unit benefit is defined (Truong-Huu, 2014) as $α_{ik}$ in $\boldsymbol{α_i}$=<$α_{i1}, α_{i2},…, α_{iR_i}$>.

Furthermore, provider $i$ have $L_i$ numbers of instances of the applications; each application $j$ has an individual per unit benefit for the provider as $β_{ij}$ in $\boldsymbol{β_i}$=<$β_{i1}, β_{i2},…, β_{iL_i}$>. The applications can be multi-tenant; a single instance of a multi-tenant application serves multiple users. Although multi-tenant applications are more expensive, providers try to increase these applications as multi-tenancy can be economical; they have greater benefit for the providers as software development and maintenance costs are shared. The instances list of applications which provider $i$ owns is $\boldsymbol{App_i}$=<$App_{i1}, App_{i2},…, App_{iL_i}$>; application $App_{ij}$={$AppID_{ij}, μ_{ij}, \boldsymbol{Srv_{ij}}, MT_{ij}, θ_{ij}$}; where, $AppID_{ij}$ presents the application $j$'s identification in provider $i$; this application is assumed to consist of $μ_{ij}$ services, $\boldsymbol{Srv_{ij}}$ is the list of services in the application in form of <$VMM_{j1}, VMM_{j2}, …, VMM_{jμ_{ij}}$>. Each service demands an individual VM type. It is to be noted that $μ_{ij}$ and $\boldsymbol{Srv_{ij}}$ are dependent on provider which hosts the application; therefore, they are determined by each provider independently. $MT_{ij}$ denotes the number of users able to own the application simultaneously; it is to be noted that $MT_{ij}>1$ for multi-tenant applications. $θ_{ij}$ is the initial price of $App_{ij}$ which is determined by its developer.

Suppose $Req_r$ demands for $AppID_{ij}$; the requested application is $App_{ij}$. The cost that SaaS provider $i$ has to pay IaaS provider for hosting $App_{ij}$ is computed as

$$c_{ij} = τ_r \times \sum_{k=1}^{μ_{ij}} VMM_{jk}.HourCost, \qquad \forall k \in [1,…, μ_{ij}]. \qquad (1)$$

After receiving a request, the provider computes the cost of processing the request. In the competition between SaaS providers, one may win due to its offered price while others lose.

*2.3 Pricing models for the applications*

Software products and requests have different properties, which affect the pricing strategies; these properties are extracted (Narahari, 2005; Lehmann, 2009; Mathew, 2010) as follows:

- **Initial cost**: the amount of money that the service provider spends for buying the software; this factor consists of the costs of the components of a SaaS based service.
- **Resource appropriation**: the efficient allocation of resources will help in reducing the wastage of resources and help keep the service as lean as possible. Eq.1 computes the resource appropriation of an application.
- **Multi-tenancy**: Number of users accessing the application simultaneously, which helps in lowering costs for the users and providers; SaaS providers can fully exploit the underlying technology.
- **User willingness to pay**: users determine an amount of money that they intend to pay for the application according to its realized value; providers do not have any knowledge of this factor, therefore, users determine it in the request.
- **Performance**: the SaaS provider guarantees a certain performance level for a determined price and pays a penalty in the case that the performance is not achieved.
- **Structure of payment flow**: users can make a single payment and thus obtain perpetual rights of use for the service, or can make a regularly recurring payment.

In this research, based on the values of these factors, different levels of services are determined; the levels affect the pricing strategy introduced in Section3. For detailed parameters that lead to dynamic pricing, we refer reader to (Narahari, 2005; Lehmann, 2009; Mathew, 2010). Different values of these parameters comprise different states, which are introduced as service levels; Table2 depicts some of the states. Users can view the details of each level while requesting in the web page of *Cloud Committee*. The values of each parameter are assigned as follows. The value of 1, for Utilization parameter, denotes that the resource appropriation of current application is the same as its requirements; false value for Multi-tenancy parameter indicates that the application is not a multi-tenant one, and true value shows a multi-tenant application is available. As mentioned previously, a multi-tenant application has a higher initial price, but as the deployment and maintenance costs are shared, users have a lower final price; Performance parameter is true, when the provider guarantees a certain performance level and determines a penalty of violation for the service; otherwise, it is false. It is to be noted that when Utilization is less than 1, Performance cannot be guaranteed and it is false. Finally, values of Payment flow parameter can be single or recurring. Services have different parameters which provide different levels of service (see Table2). Users can determine which level to access while requesting; for instance, if a user demands a typical application (not multi-tenant) with performance guarantees and single payment, then the level of service is one.

Service levels are used in pricing strategies of providers introduced in Section3.3. The initial cost and resource appropriation parameters directly affect the offered price of requests; the remaining parameters affect the price by determining different service levels. Each level individually influences the offered prices. In addition to these factors, a service provider must consider the offered prices of other service providers as well. Competition-based pricing, which sets the prices based on the other competitors' prices, is a potential dynamic pricing model. A dynamic pricing model of applications is proposed within this research by the aim of the game theory.

**Table 2** The information of service levels and the corresponding parameters of service

| service level | Utilization | Multi-tenancy | Performance | payment flow | $\omega_i$ |
|---|---|---|---|---|---|
| Level 1 | $\geq 1$ | False | True | Single | [0.1,0.15,...,0.4] |
| Level 2 | $\geq 1$ | False | True | Recurring | [0.04,0.05,...,0.09] |
| Level 3 | <1 | True | False | Recurring | [0.006,0.009,..., 0.03] |
| Level 4 | <1 | True | False | Single | [0.001,0.002,..., 0.005] |

## 3. Dynamic Pricing of Application Requests in a Competition-based Cloud Marketplace

This section firstly, studies the formulation of our proposed approach for SaaS providers' pricing model with the aim of optimizing their profit. Then, in order to establish a competition-based pricing model the setup of a game between SaaS providers is discussed.

### 3.1 Proposed Architecture

The overall architecture of the considered Cloud computing market (Fig.1) is depicted in Fig.2 with several SaaS providers. *Request Interface*, which is placed on top of the architecture, is an interface unit for received requests, **REQ**; it maps request *r* into the introduced form in Section2.1 as *Req$_r$*. The next unit, *Request Handler*, consists of two modules: *Provisioning* module and *Pricing* module. These modules have main role in processes of SaaS providers; *Provisioning* module allocates the proper available VMs, stored in *Virtual Resources* unit of the provider; *Pricing* module determines a dynamic price for the current request, $Req_r$. API sends $Req_r$ to SaaS providers, who are registered in the considered committee. The registered SaaS providers compete for serving $Req_r$. They perform allocation of resources and compute a pricing process; finally, SaaS provider $i$ sends its offer to *Market Manager* in form of $A_i$.

*Market Manager* receives offers of SaaS providers, and stores them in a vector named **s**. It is sent to the user of $Req_r$, in order to the most proper offer be chosen. *Market Manager* resends the overall information of the offers to the providers to inform the winner provider of the competition; this information is sent as a vector named **$Rep_r$**={*winner_id*, **s**}. As mentioned previous, application $j$ of provider $i$, $App_{ij}$, consists of $\mu_{ij}$ variant services. The requirements of service $m$ are specified in terms of parameters of VM, $VMM_{jm}$ in $Srv_{ij}$. In our model, goal of SaaS providers is to find the most proper price, while satisfying the requirements of the application to increase their profit. In next section, the formulation of SaaS providers for achieving the goal is studied.

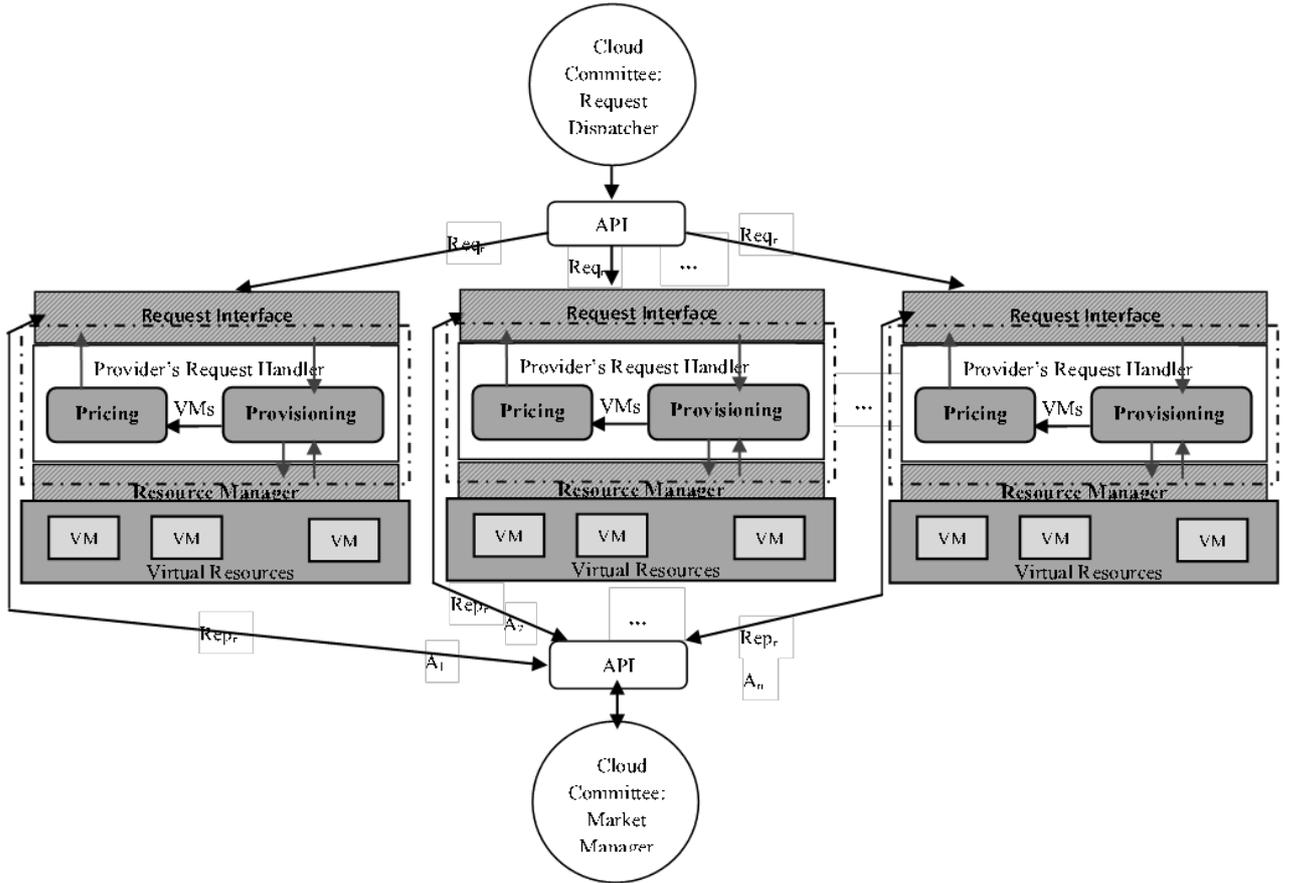

**Fig. 2** The target Cloud computing marketplace model with the considered structure of a SaaS provider

*3.2 Formulation of providers' strategies optimization*

It is supposed that SaaS providers face an optimization problem for maximizing their profits, while satisfying users. The profit of a SaaS provider is the difference between its earned revenue of processing the requests and its paid cost for providing applications and deploying them on its virtual resources. The optimization problem of SaaS provider $i$ is formulated as follows.

$$\begin{aligned}
max \ u_i &= max(P_i - C_i) = max \ P_i - min \ C_i \\
s.t \quad & P_i \leq W_r \\
& P_i \leq c_{ij} + \theta_{ij} \\
& P_i > 0, C_i > 0,
\end{aligned} \quad (2)$$

where $u_i$ is the profit of SaaS provider $i$; $P_i$ and $C_i$ are the revenue and the cost of SaaS provider $i$ while provisioning $Req_r$, respectively; $\theta_{ij}$ denotes the initial cost that provider $i$ has paid for owning the requested application, i.e. development cost of the application $j$; $c_{ij}$ is the resource appropriation cost of the requested application $j$ in provider $i$ (Eq.1).

The constraints of Eq.2 are considered to guarantee some features as follows. The first constraint is that the offered price ($P_i$) should not exceed the users' willingness to pay. It is supposed that the users' willingness to pay cannot exceed the initial cost of the application $j$ in provider $i$ ($\theta_{ij}$) and its deploying cost ($c_{ij}$) (Narahari, 2005), i.e., $W_r \leq c_{ij}+\theta_{ij}$. This assumption is applied by the committee to prevent users having low $W_r$. Second constraint shows that the offered price does not exceed the sum of $c_{ij}$ and $\theta_{ij}$; otherwise users prefer to buy the required application of $Req_r$ and its infrastructural requirements individually. The first two constraints result $P_i \leq W_r \leq c_{ij}+\theta_{ij}$. The last constraint denotes both the revenue and the cost have positive values.

A recommended solution to reach the goal is to maximize $P_i$ while minimizing $C_i$, in such a way that the constraints are maintained; the ultimate value of $P_i$, which satisfies the constraints, is achieved by some parameters, which will be discussed later.

### 3.3. A game-theoretic setup

The interaction of SaaS providers can be modeled in form of a game. Hereafter, we formulate the game for the application pricing problem in the considered Cloud computing marketplace.

*Definition1:* Let DPG=($N$,$S$,$u$) be a non-cooperative finite dynamic pricing game with complete information. $N$ is a finite set of $n$ SaaS providers in Cloud marketplace indexed by $i$; $S=S_1\times\cdots\times S_n$, where $S_i$ is a finite set of strategies of provider $i$, which presents its pricing policies; $u=(u_1,...,u_n)$, where $u_i$ is payoff function of provider $i$. Let $s=(s_1, ..., s_n)\in S$ as the strategy profile, where $s_i\in S_i$ is the strategy of player $i$; $s_i$ is chosen in a way to maximize $u_i(s)$.

Players of DPG (dynamic pricing game) are a set of SaaS providers of Cloud computing which registered in *Cloud Committee*. Although the number of SaaS providers are rapidly grows there are a finite number of providers in Cloud environment (Hurwitz, 2010); thus, we have a finite set of players, which is a necessity in a finite game. Users can easily find the latest list of SaaS providers that offer software solutions in their interested area. SaaS providers who register in *Cloud Committee* have a common database which makes DPG a complete information game.

SaaS provider can discriminate the prices (Narahari, 2005; Lehmann, 2009) based on the per unit benefits that each application and VM have; the price discrimination offers a same application to different users at different prices, based on the introduced factors in Section2.3.

We use a competition-based pricing model, realized by designing the considered non-cooperative game, which has also benefited from price discrimination. $S_i$ denotes the strategy set of provider $i$ as,

$$S_i = \sqrt{\omega_i}(1 + \gamma\sqrt{\omega_i})(\theta_{ij} + c_{ij}), \tag{3}$$

where, $\omega_i$ is a parameter which is determined by provider $i$, $\gamma$ is a constant value determined by the committee, less than 1.

Let $\rho_r$ denote Utilization parameter for $Req_r$, which demands for $AppID_{ij}$, as

$$\rho_r = \frac{R_r}{R_i}. \tag{4}$$

Where $R_r$ is the required infrastructures of requested application in $Req_r$, $R_i$ denotes the provided resources for the request by provider $i$. The greater values of $\rho_r$ guarantee a certain performance level for a determined price; otherwise, for values of $\rho_r$ less than 1, i.e. ($R_r < R_i$), the offered price is discounted.

The other factors that influence $\omega_i$ are as follows; multi-tenancy state of the requested application, the user's interest for performance guarantees and payment flow, which are discussed in Section2.3. These factors determine different service levels. We determine different range of $\omega_i$ according to each service level. Let $C_\rho$, $C_{MT}$, $C_{Pay}$ and $C_{Prf}$ denote Utilization, Multi-tenancy, Payment flow, and Performance, respectively. The service levels are determined according to these parameters (see Table2); $\omega_i$ has different ranges of values based on each service level. Lower service levels have higher prices; i.e., for initial levels of service, high values of $\omega_i$ (e.g. [0.1,0.15,...,0.4]) are considered in order to not decrease the price of the provisioned request; these levels would have smaller discounted prices, and vice versa; the supposed ranges of $\omega_i$ are discussed in Section5.1.

Service provisioning is a cost-prone process; however, there is a trade-off between revenues and costs. Formally, payoff function, which introduces the profit of a provider, consists of both the properties of users' demand in $Req_r$, and the corresponding properties of the provided service.

*Definition2:* For the strategy profile $s\in S$, $u_i:S\rightarrow\mathbb{R}$ is the payoff function, which assigns numerical values to each member of the strategy set $S$. For all $x,y\in S$, strategy $x$ is preferred over strategy $y$ iff $u_i(x)>u_i(y)$.

A payoff function is originated from software pricing principles discussed in Section2.1 (Narahari, 2005; Lehmann, 2009; Mathew, 2010). The payoff function of player $i$, for $Req_r$, is

$$u_i(s) = D_i(S_i - C_i), \qquad (5)$$

where $S_i$ and $C_i$ are the strategy of provider $i$ for providing the request, respectively. $D=\{D_1, D_2, ..., D_n\}$ denotes the demand vector of SaaS providers; $D_i=1$ if and only if $argmin(s)=i$, i.e. $i$ is the index of the minimum of profile $s$ and the strategy of player $i$ has the least value in profile $s$; therefore player $i$ wins the game, otherwise it is zero. $u_i(s)$ not only depends on the strategy of provider $i$ but also depends on all others', i.e. $s=(s_1, s_2, ..., s_n)$. Users usually choose the least price for a service with a satisfactory performance; therefore, the payoff is zero unless for the provider with the least price. $C_i$ is computed as,

$$C_i = \omega_i(\alpha_j c_{ij} + \beta_{ij}\theta_{ij}). \qquad (6)$$

$\omega_i$ is determined by each provider individually; it is applied to ensure the positivity of $u_i$. $\beta_{ij}$ and $\alpha_j$ are per unit benefits of application $j$ for provider $i$ and per unit benefits of virtual resources that host $Req_r$, respectively; $\alpha_j = \sum_{k=1}^{\mu_{ij}} \alpha_{ik}$, where, $\alpha_{ik}$ denotes the per unit benefits of VM $k$ for provider $i$.

Finally, the payoff function represented in Eq.5, is simplified as

$$u_i = D_i\left(\sqrt{\omega_i}(1 + \gamma\sqrt{\omega_i})(\theta_{ij} + c_{ij}) - \omega_i\left((\sum_{k=1}^{\mu_{ij}} \alpha_{jk})c_{ij} + \beta_{ij}\theta_{ij}\right)\right). \qquad (7)$$

The strategy profiles must converge to a desired profile, which is known as the solution concept of the game. This solution is named as Nash equilibrium in a normal form game; next section investigates the equilibrium.

Algorithm1 presents the algorithm for DPG. The output is the list of providers' pricing offers in Nash equilibrium. Firstly, the provider receives a request from *Request Dispatcher* of *Cloud Committee* in line1. Then, the game starts and it proceeds until the equilibrium achieved. Providers have some virtual resources for deploying the requested application in lines3 and 4. Each provider specifies a value to $\omega$ via line5 (Table2); the price offer is computed in line6 (Eq.3). The offered values of all providers are saved in *BidList*, which is a distributed memory between providers and the *Market Manager* of *Cloud Committee*. In line7, the winner of the game is found. The payoffs are computed through Eq.7 in lines8-11. Finally, achieving Nash equilibrium is checked in lines12-13; the process checks, whether any of the providers can get more benefit by changing the current offer, while other providers do not change their strategies.

**Algorithm 1** resource provisioning Game algorithm
The algorithm is run by each SaaS provider of the committee as CurrentPrv in a distributed manner.
**Input**:
Request of applications, $Req_r$; Information of SaaS providers: list of VMs and list of applications, their benefits $\alpha$,
$\beta$, $\gamma$, $D$=0;
**Output**:
Optimal list of prices, BidList

```
1   Req_r = CloudCommittee.Dispatch();
2   Do
3      foreach Service in associated AppID of Req_r do
4         SelectedVMList[Service] = Select a proper VM for Service;
5      ω= Selectω(Req_r);
6      BidList[CurrentPrv] = S_CurrentPrv (Req_r, θ_AppID, SelectedVMList, ω, γ);
7      Winner = MarketMgr(BidList);
8      If CurrentPrv matches Winner
9         Cost = C_CurrentPrv(Req_r, SelectedVMList, α_CurrentPrv, β_CurrentPrv);
10        D_CurrentPrv=1;
11        u_CurrentPrv = BidList[CurrentPrv] - Cost;
12     If MarketMgr() matches NE then
13        Return BidList;
14  While (1)
```

The complexity of this algorithm is $n \times |S|$, where $n$ is the number of available SaaS providers and $|S|$ is the size of the strategy set of the provider.

Next section investigates the properties of Nash equilibrium for the game.

## 4. Market Equilibrium

Cloud computing is a complex and heterogeneous distributed environment, in which management of the interactions between entities is a challenging task and needs automated and integrated intelligent strategies. States of SaaS providers in Cloud computing environment are unpredictable; therefore, predicting the behavior of providers accurately, would be a costly task. For this reason, we applied game theory concepts to simplify the problem of dynamic pricing of applications in Cloud computing market; in this section, the properties, existence, and uniqueness of solution concept of DPG are discussed.

### 4.1 Nash equilibrium conditions for the game

Unfortunately, the problem of finding Nash equilibrium of a general-sum game with *n* players cannot be formulated as a linear program (Shoham, 2008); thus, we cannot state the problem as an optimization problem as presented in Eq.2.

In DPG, providers determine pricing strategies, which satisfy them with their expected payoff, known as Nash equilibrium (Fudenberg, 1996). So, Nash equilibrium is an optimal criterion for DPG, which none of the SaaS providers can get more benefits by changing the selected strategy unilaterally; in Nash equilibrium, the assumption is that the other providers do not change their strategies.

Previously mentioned, the strategy of players, $S_i$, is a linear function of parameters related to the request and application; e.g., initial price of application, resources appropriation, and their benefit list. Each SaaS provider chooses the value of $\omega_i$ from a predefined range of finite values determined based on level of provided service (Table2). As presented in Algorithm1, they will continue choosing these values until the equilibrium condition is satisfied.

Hereafter, the existence of at least equilibrium and its uniqueness will be studied.

### 4.2 Nash equilibrium existence and uniqueness

The termination condition of Algorithm1 in Section3.3 is to achieve Nash equilibrium; in Theorem1 it is proven and discussed.

**Theorem1** There is at least one Nash equilibrium for DPG.
**Proof** Shoham (Shoham, 2008) has proven that every game with a finite number of players and a finite number of strategies has at least one Nash equilibrium. DPG has a finite number of players, which are SaaS providers in Cloud environment (Buyya, 2009). Besides, both strategy profile and payoff function of DPG are finite, as their parameters have finite values. Therefore, Shoham's theorem verifies the existence of Nash equilibrium in DPG. If the values of these parameters were chosen from a continuous value set, then catching Nash equilibrium in Algorithm1 (line12) would have the complexity of $n^n$; therefore, some other intelligent strategies would be needed.

Finally, in order to guarantee the termination of the game, the uniqueness of Nash equilibrium is discussed as well.

**Theorem2** DPG has a unique Nash equilibrium.
**Proof** Based on Theorem1 and the well-known Weierstrass theorem (De Branges, 1959), $u_i$ is a closed function as it is a finite function (Rudin, 1964). The Weierstrass theorem guarantees that every function defined on a closed interval can be uniformly approximated by a polynomial function. This polynomial function can be assumed a linear function. $u_i(s)$ consists of several polynomial terms, which are linear. The concavity of $u_i(s)$ can be easily proved by studying its linear terms; as $ax+b$ can be supposed as a concave function, $S_i$ is a concave one as well. On the other hand, $C_i$ is an affine too, and it is concave. Consequently $u_i$, which is $S_i$-$C_i$ is a concave function on convex set $\omega_i$. It is to be noted that $u_i(s)$ is a second-order differentiable and concave function of its parameters (Chen, 2011), which guarantees the convergence of DPG.

Based on the concavity of $u_i(s)$, an equilibrium point, $s^o$, of a game with a concave payoff function can be assumed as the following.

$$u_i(s^o) = max_{y_i}\{u_i(s_1^o, \cdots, y_i, \cdots, s_n^o) | (s_1^o, \cdots, y_i, \cdots, s_n^o) \epsilon S\} \quad (i = 1, \cdots, n) \qquad (8)$$

At point $s^o$, every provider stays in its best state and never changes the strategy while other strategies are unchanged. After considering the fact that DPG is a concave game with *n* players, the uniqueness of Nash equilibrium is proved by using standard techniques based on (Rosen, 1965).

Based on Theorems 1 and 2, DPG would have an individual Nash which is known as the solution concept. Thus, the game finds a solution for providers to reach the most available profit; in next section, this solution in a duopoly is discussed and the strategy of players in Nash is presented in form of a closed-form in duopoly.

*4.3 Closed-form expression of the pricing strategy*

In this section, Nash equilibrium of DPG in a duopoly is studied; the proof of a duopoly can be generalized to a scenario having more than two SaaS providers. Nash equilibrium price can be obtained through the best response function of each player in a non-cooperative game (Chen, 2011); i.e. $s^*$ is considered as Nash equilibrium if $s^*_i$ is the best response of provider $i$:

$$\begin{aligned} u_1(s^*) = u_1(s_1^*, s_2^*) \geq u_1(s_1, s_2^*) &, \forall s_1 \epsilon S_1 \\ u_2(s^*) = u_2(s_1^*, s_2^*) \geq u_2(s_1^*, s_2) &, \forall s_2 \epsilon S_2 \end{aligned} \quad (9)$$

The optimal $s$ corresponding to maximal $u_i(s)$, which is the best response of provider $i$, is computed by differentiating $u_i(s)$ with respect to $s$; then, it is set to zero, as follows

$$\frac{\partial u_i}{\partial s} = D_i \left(\frac{1}{2\sqrt{\omega_i}} + \gamma\right)\left(\theta_{ij} + c_{ij}\right) - D_i\left(\left(\sum_{k=1}^{\mu_{ij}} \alpha_{jk}\right)c_{ij} + \beta_{ij}\theta_{ij}\right). \quad (10)$$

Jointly solving the expression $\frac{\partial u_i}{\partial s} = 0$, the optimal pricing policy of provider $i$ in a duopoly can be obtained as the closed-form expression of pricing policies. With a view to the parameters of $s_i$ consists of $\gamma$, $\theta_{ij}$, $c_{ij}$, and $\omega_i$, $\gamma$ is a constant coefficient determined by *Cloud Committee* marketplace, $\theta_{ij}$ is a value defined by the developer of the application, $c_{ij}$ is computed based on resource appropriations, and $\omega_i$, which is determined by provider $i$, equals to the following value in equilibrium point,

$$\omega_i^* = \left(\frac{\theta_{ij} + c_{ij}}{2\left(\left(\sum_{k=1}^{\mu_{ij}} \alpha_{jk}c_{ij} + \beta_{ij}\theta_{ij}\right) - \gamma(\theta_{ij} + c_{ij})\right)}\right)^2. \quad (12)$$

The best response of each player in the considered duopoly is $s = (\sqrt{\omega_1^*}(1 + \gamma\sqrt{\omega_1^*})(\theta_{1j} + c_{1j}), \sqrt{\omega_2^*}(1 + \gamma\sqrt{\omega_2^*})(\theta_{2j} + c_{2j}))$. Consequently, closed-form expression of pricing policies of provider $i$ in a duopoly is $\sqrt{\omega_i^*}(1 + \gamma\sqrt{\omega_i^*})(\theta_{ij} + c_{ij})$, which is known as the solution of duopoly market in DPG.

**Table 3** Pricing defined by IaaS provider

| Attr. Size | VCPU | Memory (GB) | Storage (GB) | Price per VM/$ |
|---|---|---|---|---|
| t2.small | 1 | 2 | 1x 4 SSD | $0.026/Hour |
| t2.medium | 2 | 4 | 1x 4 SSD | $0.052/Hour |
| m3.medium | 1 | 3.75 | 1x 4 SSD | $0.070/Hour |
| c3.large | 2 | 3.75 | 2x 16 SSD | $0.105/Hour |
| m3.large | 2 | 7.5 | 1x 32 SSD | $0.140/Hour |
| R3.large | 2 | 15 | 1x 32 SSD | $0.175/Hour |

## 5. Performance Evaluation

In this section, some experiments for analyzing the proposed model of competition of SaaS providers in Cloud computing marketplace are developed. Firstly, the parameter settings and performance metrics are studied; then, the simulation configurations are explained, and finally the results are presented.

*5.1 Experimental Setup*

In this section, the parameters and the configuration of DPG are clarified. The experiments are run on a semi Cloud computing marketplace, CloudSim toolkit 3.0.2, as follows.

*5.1.1 Parameters setting*

The considered marketplace consists of multiple SaaS Cloud providers, which initially owned random number of different type of VMs. Three methods are added to implement the process of SaaS providers in the market. The first method is used to determine whether the provider can provide the received request based on its virtual resources or not. This method investigates the properties of available VMs by the aim of providers' *Resource Manager* and the requirements of each request by the aim of *Request Interface*. For simplicity, a single service is deployed on each VM. The second method finds the most proper VMs for deploying the requested application. This method chooses a VM, which is capable of deploying the service in a low per hour cost, for all services in the application. The third method specifies a price for supporting the request.

The parameters of VMs such as size, memory, and price are considered based on what Amazon EC2 has defined (in December 2015). The parameters and the prices of VMs are listed in Table3. In the experiments, each VM hosts just one service. *Vm* class of CloudSim toolkit is extended to support the mentioned properties of *VMM* in Section2.1, based on Table3.

In the experiments, $\gamma$ is set to 0.95 (Truong-Huu, 2014); it corresponds to a 0.05 interest rate. The parameter $\omega_i$ is derived from a finite set based on the level of the service. The probability distribution of values of $\omega_i$ is initialized as uniform distribution.

*5.1.2 Simulation Configuration*

Cloud environment is modeled in form of one IaaS provider, several SaaS providers and some users. In the simulations, we assumed two or 10 number of SaaS Cloud providers with a single IaaS provider and different provided VMs.

The requests of users are modeled as application demands in form of $Req_r$. These requests include execution-related requirements of applications such as memory, CPU usage, etc.

Moreover, each SaaS provider in our supposed Cloud computing marketplace owns multiple applications, and each application may consist of several services; the same list of ERP applications (Enterprise Resource Planning) is considered in each SaaS provider. Different ERP applications are provided from different SaaS providers; CRM is an ERP application, which may have three main instances: Essential, Basic, and Professional. Some instances of Microsoft CRM applications and their potential costs are presented in Table4. Cloud applications' costs vary based on commercial fees[1]. Providers are monthly billed per user for online provisioning; the licensing prices are determined based on the instances, for on premise provisioning. The prices of our applications are derived from ERP providers, such as Actionstep, iCIMS, Plex Systems and Host Analytics Inc.; the assumed values of parameters of the simulation are considered like (Truong-Huu, 2014)[2].

**Table 4** Considered applications offered by SaaS providers with their costs[3]

| Type of Provision<br>Application's License | On-Premise | Online<br>(per user per month) |
|---|---|---|
| CRM Server 2013 | $4922 | $150 |
| CRM Professional User CAL | $983 | $65 |
| CRM Professional Device CAL | $787 | $65 |
| CRM Basic User CAL | $342 | $30 |
| CRM Basic Device CAL | $236 | $30 |
| CRM Essential CAL | $79 | $15 |

*5.2 Equilibrium Efficiency*

To validate the correctness of our proposed competition-based pricing approach, we run the experiments in a duopoly market with two SaaS providers. In the following experiments the unit of profit and prices are $ and iteration denotes the number of repetitions which has not any unit. By such an assumption, we simplified the experiments, while retaining the competitive characteristics of the considered market. Fig.3a shows the profit of two providers while receiving different requests. As depicted in this figure, while the game proceeds, both providers almost obtain an increasing profit.

Then, the experiments are performed for more than two providers to validate the approach in an oligopoly Cloud market. Fig.3b shows the profit of SaaS providers ($N$=10). From Fig.3b, it can be observed that while the game proceeds, the profits mostly increase as well. These experiments assess the performance of the competitive pricing mechanism. As it can be observed in Fig.3, the sum of profits of providers in duopoly and oligopoly with the same conditions are approximately equal to each other. One reason for large differences of profits of providers as depicted in Fig.3b is that these SaaS providers have different **α** and **β** (per unit benefits of resources and applications, respectively).

---

[1] https://www.softwareadvice.com/crm/
[2] https://aws.amazon.com/getting-started/
[3] https://www.softwareadvice.com/crm/

Fig.4 depicts the evolution of offered prices of providers, which is inserted as bids. Provider $i$ chooses pricing parameter, $\omega_i$, randomly, that turns out different prices. In each iteration, Nash equilibrium circumstances, introduced in Section4.1, are checked. As shown in Fig.4a, in 17$^{th}$ iteration, both providers reach Nash equilibrium, where their profit is better than their other offers. As depicted in the figure, the offers of providers are not changed after reaching Nash equilibrium; this state is called the convergence point of the game.

Fig.4b repeats the experiments for $N$=10 providers. In some cases, the game runs more than 100 iterations to reach the equilibrium. Providers get Nash equilibrium in iteration 43$^{rd}$, where their profit is better than other offers, while the other providers do not change their offers.

Comparing Fig.4a and Fig.4b, it is to be noted that in a multiple-players game, the convergence of pricing policy in an oligopoly needs longer run, which is expected as the growth of strategy profile of players. The simulation results verify that the considered game always converges to the optimal solution known as Nash Equilibrium. Actually, price offering of providers converges to the optimal price. The optimal price is the least one that satisfies the constraints in Eq.2. In Fig.4a it can be obviously observed that provider 2 is the winner of the game.

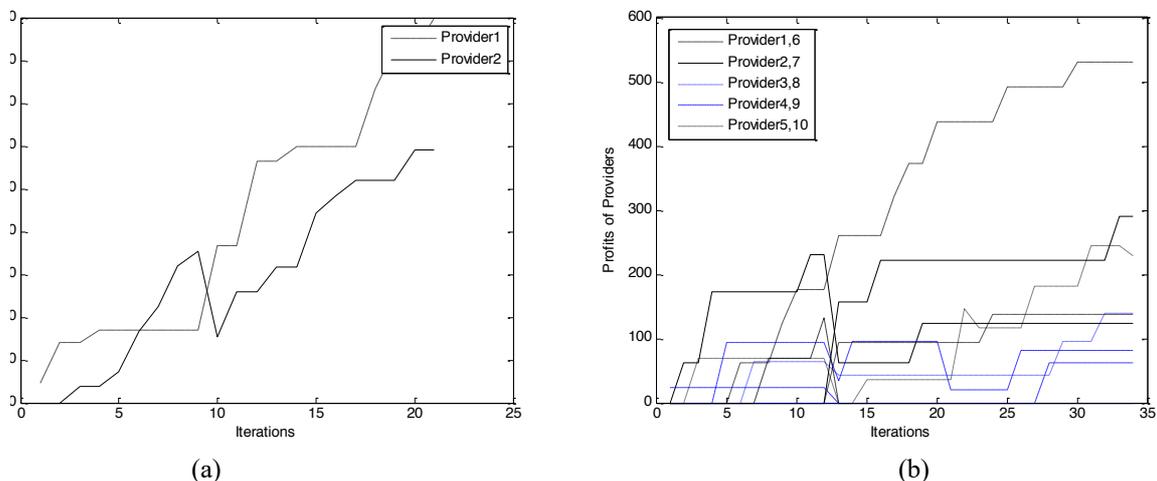

(a)            (b)

**Fig. 3** Payoff of providers in a (a) duopoly, (b) oligopoly Cloud computing market

The states of providers in Nash equilibrium are depicted in Table5. It is to be noted that a demanding request of service level 1 is supposed; based on the same price of application ($\theta$), and resource appropriation costs ($c$), different per unit benefits, different strategies ($s_i$) are generated. The winner is provider $i$=9; therefore, the profit of all players except 9 is actually zero; however, as depicted in Table5, their imaginary profits, in case they are the winner, are presented to enable the comparison.

**Table 5** Presentation of providers' strategies parameters while $\gamma$=0.95, $\theta$=65$, $c$=295$

| Provider $i$ | $\alpha_i$ | $\beta_i$ | $\omega_i$ | $s_i$ | Cost$_i$ | Profit$_i$ |
|---|---|---|---|---|---|---|
| 1 | 0.1 | 0.4 | 0.001 | 11.20504 | 0.0599 | 11.14514 |
| 2 | 0.3 | 0.25 | 0.001 | 11.56334 | 0.0975 | 11.45984 |
| 3 | 0.8 | 0.25 | 0.001 | 11.20504 | 0.23945 | 10.96559 |
| 4 | 0.4 | 0.5 | 0.001 | 11.7262 | 0.1505 | 11.5757 |
| 5 | 0.2 | 0.4 | 0.001 | 11.23761 | 0.081 | 11.15661 |
| 6 | 0.8 | 0.15 | 0.001 | 12.21479 | 0.25125 | 11.96354 |
| 7 | 0.1 | 0.7 | 0.001 | 12.37766 | 0.077 | 12.30066 |
| 8 | 0.4 | 0.5 | 0.001 | 11.7262 | 0.1505 | 11.5757 |
| 9 | 0.2 | 0.15 | 0.001 | 11.0096 | 0.1026 | 10.94549 |
| 10 | 0.3 | 0.7 | 0.001 | 11.56334 | 0.10325 | 11.43084 |

**Table 6** The length evolution of game while the players of the game increase

| Criterion / Approach | Duopoly | | Oligopoly | |
|---|---|---|---|---|
| | average of best price | average of profit | average of best price | average of profit |
| DPG | 12.43 | 11.21 | 11.2 | 10.48 |
| Price discovery | 15 | 11 | 14.3 | 10.5 |

Finally, some experiments are performed to compare the approach with other methods to identify the breakthrough that has been achieved using DPG model. In (Muzaffar, 2017), a price discovery algorithm for searching the optimal price for a service with price-sensitive demand is studied which no information is required on reservation price. Our approach is compared with (Muzaffar, 2017); the average of the best price and the profit of providers are considered in the comparison. The same demand rate of the applications is assumed in experiments. The results are depicted in Table 6, as follows.

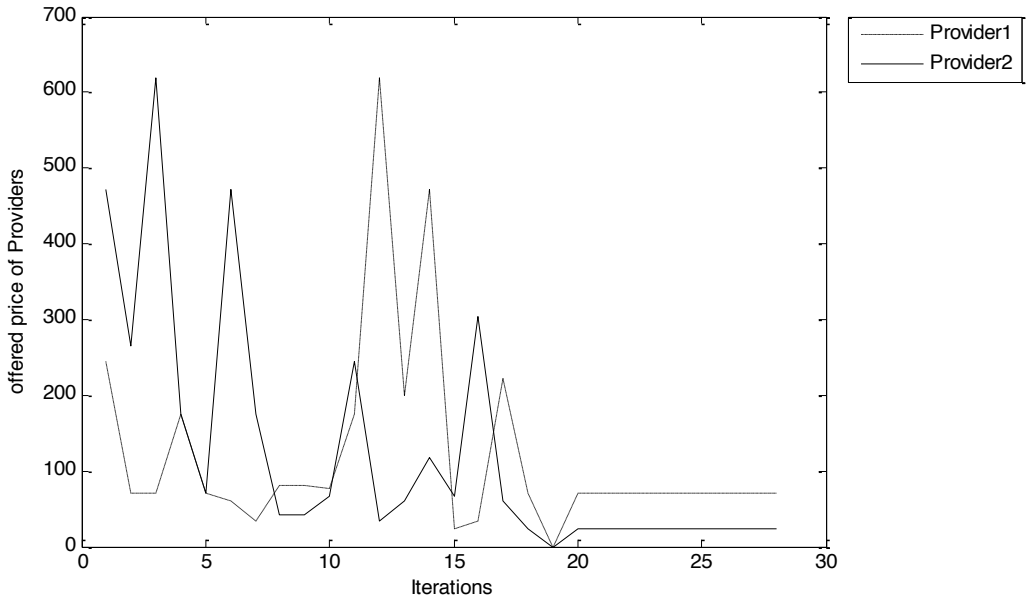

(a)

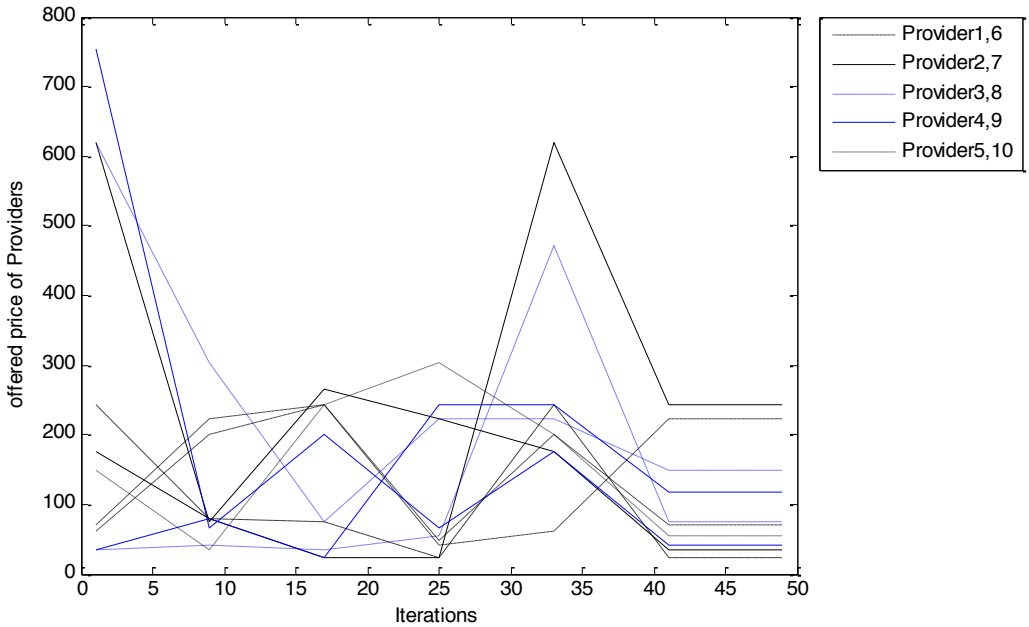

(b)

**Fig. 4** Evolutions of SaaS providers' bids in a (a) duopoly, (b) oligopoly (N=10) Cloud computing market to equilibrium state

DPG and Price discovery approach of (Muzaffar, 2017) are performed in both duopoly and oligopoly markets which have two and 10 SaaS providers. They have multiple instances of Microsoft CRM applications whose costs are presented in Table4. As depicted in Table6, in both markets while using DPG the best prices are less than (Muzaffar, 2017). Although the offered prices in our approach is less the average of profits that providers gain is approximately the same in both approaches. The reason is that the number of requests that each provider may serve increases when it wins the game.

*5.3 Validating the scalability of algorithm*

Previous experiments considered at most $N$=10 SaaS providers in oligopoly markets; however, the proposed algorithm can scale to a realistic size of players without disobeying time limits. On a Macbook Core 2 Duo running at 2.40GHz with 4.0GB RAM, the number of players is exponentially increased, with different requests and parameters. It is expected that as the number of players grows the execution time of the algorithm increases as well; it can be also concluded from Fig.4. As illustrated in Table7, the affects that the growth of game's size has on the length of the algorithm's run is not exponential. Table7 depicts the average number of iterations and the average of elapsed time of the game for reaching the equilibrium in ten runs. The algorithm must check whether or not the equilibrium is achieved, in each iteration. As discussed previously, the order of Algoritm1 is $O(n^2)$. The longest time required for reaching the equilibrium, for 1024 SaaS providers, is 834 seconds.

**Table 7** The length evolution of game while the players of the game increase

| players | 2 | 8 | 16 | 32 | 64 | 128 | 256 | 524 | 1024 |
|---|---|---|---|---|---|---|---|---|---|
| Iteration to reach NE | 27 | 40 | 48 | 80 | 100 | 110 | 120 | 125 | 140 |
| Elapsed time (s) | 1.297 | 1.711 | 1.879 | 2.342 | 9.421 | 20.162 | 48.521 | 107.945 | 276.436 |

## 6. Conclusion

Recently Cloud computing has been emerged as a new information technology development, which has been noted as a services marketplace. Thid marketplace faces with the competition and cooperation of its providers; this research focuses on the competition of Cloud providers. SaaS providers compete with each other by offering suitable resource provisioning with a desirable price. The scenario is modeled in a finite normal form game. Players of the game are SaaS providers; their strategies are considered as competition-based dynamic pricing policies based on different application properties; their preferences are the revenue, which is obtained by providing the request with offered price. We verified the existence and uniqueness of Nash equilibrium for the game. In addition, the experimental evaluations are performed and the theoretical evaluations are verified. Providers seek equilibria to perform an adaptive pricing strategy; the considered game, which computes the preferred dynamic prices for each provider, converges to a unique Nash equilibrium, in which none of the providers tend to change their strategies.

Assumption of having just one IaaS provider can be omitted in the case of extending the model in the future to focus on resource provisioning techniques of providers. The infinite set of strategies is another issue for our future study.